# Which Humans? Inclusivity and Representation in Human-Centered AI

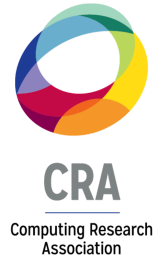 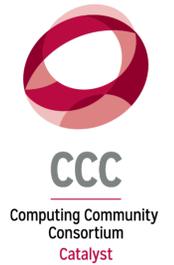

Rada Mihalcea (University of Michigan), Nazanin Andalibi (University of Michigan), David Jensen (University of Massachusetts Amherst), Matthew Turk (Toyota Technological Institute at Chicago), Pamela Wisniewski (Vanderbilt University), Holly Yanco (University of Massachusetts Lowell)

As AI systems continue to spread and become integrated into many aspects of society, the concept of "human-centered AI" has gained increasing prominence, raising the critical question of which humans are the AI systems to be centered around.

## Introduction

As artificial intelligence (AI) systems continue to be integrated into many aspects of society, spanning domains such as healthcare, education, agriculture, and transportation, the concept of "human-centered AI" has gained increasing prominence. Human-centered AI involves placing humans at the center of every phase of AI design and deployment. This includes identifying societal problems that are suitable for AI-enabled solutions, collecting real-world training datasets, and using AI techniques that do not lead to harms such as discrimination. Human-centered AI also emphasizes transparency, explainability, and adaptability to different human contexts, and it places emphasis on evaluations that ensure the AI systems benefit those directly affected, while also identifying and addressing any negative consequences before widespread adoption.

Yet, this raises a critical question: which humans are these systems centered around, and what are the implications of who is represented and who is not? Despite the promise of AI to enhance lives globally, many AI systems fail to address socially relevant problems or attempt to address problems that are relevant to a selected few. This is partly because the focus in AI development has been primarily from the "systems" perspective, rather than on the "humans" impacted by these systems. Further, for the systems developed to date, we have seen significant disparities in how benefits and harms are distributed across different communities.

This quadrennial paper explores the need to place the emphasis on "human" in human-centered AI. It calls for learning from fields such as Human-Computer Interaction and Science and Technology Studies, which have a long history of including humans in technology design and evaluation, and making informed decisions about when equal representation in AI development and deployment is beneficial or harmful to the impacted groups. The paper highlights opportunities for human-centered AI that shift the focus from the AI systems to the humans and considers a broad spectrum of human experiences and needs.

## Support the Development of AI Systems That Respect and Serve a Broad Spectrum of Human Experiences

Why is human-centered AI important, especially to serve and protect different groups? Human-centered design is critical to the AI development process because it can pinpoint aspects that technologists focusing on technical solutions may otherwise miss. For example, when HealthCare.gov launched in 2013, in addition to the issues of system performance, [many users found the website difficult to use](#) due to confusing navigation, a lack of explanation for complex insurance terms, and overwhelming information. A human-centered approach could have conducted user studies with a diverse applicant pool to uncover and remedy issues before deployment.

As another example, many law enforcement agencies began using facial recognition, which often [perform poorly on brown and Black faces](#), leading to wrongful arrests and other serious consequences. Similarly, AI systems that claim to infer human emotion are deployed across high-stakes contexts [such as hiring](#), without attending to potential harms that can [deny people access to employment opportunities](#). A human-centered approach could have identified existing biases in the training datasets, embedded necessary checks and balances, and interrogated and rectified issues before they caused widespread harm. Simply put, deploying AI systems that directly affect humans cannot be executed ethically without deeply understanding the needs of humans. But then we must ask: which humans?

## Make the Design of the AI Systems Participatory With the Stakeholders Groups

One answer to "which humans?" would be those who develop AI systems — namely, the computing research community, which has historically lacked broad representation across demographic groups. This underscores the importance of investing in national efforts for broadening participation in computing so that AI developers' systems are more representative of the populations impacted by these systems.

A second answer would be the end users or direct and indirect stakeholders impacted by AI. For instance, algorithmic decision-making systems are used in our child welfare system to determine child placement and identify risk outcomes. Failing to engage caseworkers, foster youth, and foster families in the design and development of these systems — and removing necessary human discretion and expertise — can lead to decisions that are detrimental to the safety and well-being of foster youth, as well as to caseworker retention. To realize the potential of AI in benefiting and not harming people, it is crucial to begin not with AI as the assumed solution, but with a human-centered understanding of the problem and a collaborative exploration of possible solutions, which may or may not involve AI.

Participatory approaches help anticipate and mitigate harms early and increase the acceptability of technologies, if and when they are deployed.

Some key challenges in developing participatory AI including:

- Reconciling system design values that may be in tension with each other
- Realizing that the desire to scale may conflict with more community-centered AI solutions.

## Support the Development of Frameworks and Guidelines to Evaluate the Impact of AI on Different Communities

Developing the ability to evaluate an AI system across the full range of people and communities it impacts is imperative to ensuring equitable benefits while preventing harm. Given the examples of biases in AI systems, it is critical that evaluation includes a large set of test cases. It could also be beneficial to examine training data to identify over- or under-represented examples — although such data is often proprietary. Guidelines for anonymizing datasets while preserving the ability to detect gaps in representation would help address this challenge.

There is a clear need for public-private partnerships to create robust frameworks and guidelines for assessing the impact of AI on different populations. These efforts must include participants from the communities most likely to be affected. Frameworks developed solely by large companies may lack transparency and could unintentionally limit smaller companies' ability to compete. For human-centered AI systems to succeed, the voices of those most impacted must be incorporated into State or Federal guidelines.

# Prioritize Socially Preferable AI Systems That Effectively Support the Long-Term Future of Humanity

AI system development should begin by identifying social problems that AI is well-suited to address. Rather than prioritizing AI solutions based on the limited perspective of a small group of developers, decisions about which problems to solve should be guided by collaboration with varied communities and stakeholders. This ensures that AI development aligns with real needs and values, producing solutions that are effective and socially responsible.

Moreover, we need to revisit the criteria on which AI systems are evaluated. Instead of implementing evaluation systems that only measure accuracy, we need a multi-faceted set of values that balance metrics like fairness, transparency, and alignment with impacted communities' values. In order to ensure that we develop systems that are truly "socially preferable," we must conduct long-term assessments that include evaluations of economic effects, environmental sustainability, and metrics related to the overall well-being of humanity. Ultimately, the goal is to create AI systems that not only solve immediate problems but also contribute to building a fairer, more sustainable world in the long term.

## Recommendations

Moving toward human-centered AI requires a multi-pronged approach involving technologists, communities, impacted individuals, and governments. Technologists should avoid prioritizing AI systems based on the limited perspective of a small group of developers, and instead engage with a diverse range of stakeholders throughout all phases of development — from ideation and design to benchmarking, model development, evaluation, and deployment. Communities should offer local insights and help ensure that design choices reflect real-world needs. Governments should establish frameworks and policies that lower barriers to participation, including:

- Consulting community partners when developing solicitations for research proposals for socially impactful AI research.
- Evaluating research proposals not only for community engagement but also for community leadership.
- Simplify infrastructure funding for partners unfamiliar with government funding systems.
- Mandating regular audits of AI systems' social impacts to ensure transparency and enable external evaluation.

*This quadrennial paper is part of a series compiled every four years by the **Computing Research Association (CRA)** and members of the computing research community to inform policymakers, community members, and the public about key research opportunities in areas of national priority. The selected topics reflect mutual interests across various subdisciplines within the computing research field. These papers explore potential research directions, challenges, and recommendations. The opinions expressed are those of the authors and CRA and do not represent the views of the organizations with which they are affiliated.*

*This material is based upon work supported by the U.S. National Science Foundation (NSF) under Grant No. 2300842. Any opinions, findings, and conclusions or recommendations expressed in this material are those of the authors and do not necessarily reflect the views of NSF.*